\documentclass[journal=aamick,manuscript=article]{achemso}

\usepackage[version=3]{mhchem} % Formula subscripts using \ce{}
\usepackage{enumitem}

\author{Ashlee M. Garc\'ia}
\email{ashgarcia@utexas.edu}
\author{Byron D. Aguilar}
\author{William J. Doyle}
\affiliation[UT Austin]
{Microelectronics Research Center and ECE Department, The University of Texas at Austin, 10100 Burnet Rd., Bldg. 160, Austin, Texas 78758, USA}

\author{Pernille Undrum Fathi}
\affiliation[harvard]{Harvard University, Applied Physics, 9 Oxford St,  Mckay 105, Cambridge MA 02138, USA}

\author{Federico Capasso}
\affiliation[harvard]{Harvard University, Applied Physics, 9 Oxford St,  Mckay 105, Cambridge MA 02138, USA}

\author{Daniel Wasserman}
\author{Seth R. Bank}
\email{sbank@ece.utexas.edu}
\affiliation[UT Austin]
{Microelectronics Research Center and ECE Department, The University of Texas at Austin, 10100 Burnet Rd., Bldg. 160, Austin, Texas 78758, USA}

\title[An \textsf{achemso} demo]
  {Surface Modification for III-V Selective Area Molecular Beam Epitaxy of Non-Selective Mask Materials}

\begin{document}


\begin{abstract}
Selective-area embedded regrowth of III–V semiconductors by molecular beam epitaxy enables the seamless integration of metals and dielectrics into crystalline material for novel design of optoelectronic devices. However, traditional masks like \ce{SiO2} and \ce{Si3N4} limit the design of high-contrast photonics in the infrared due to their high extinction coefficients at technologically relevant wavelengths. Consequently, there is a need to explore alternative mask materials to expand the selective area molecular beam epitaxy capabilities beyond those traditionally used. This study evaluates the deposition selectivity of the alternative materials \ce{Al2O3}, \ce{TiO2}, and \ce{HfO2}, films with preferable spectral responses but higher surface reactivity. It was found that \ce{Al2O3} exhibits promising selective growth characteristics within typical GaAs growth temperatures, \ce{HfO2} demonstrated a high non-selectivity dominated by Ga adsorption on the mask at temperatures up to 650 $^\circ$C, and \ce{TiO2} proved reactive during deposition. To achieve selective growth of highly non-selective and even reactive mask materials, a surface modification technique was employed to improve the selective growth characteristics of any given film. Selective growth of \ce{Si3N4} and \ce{TiO2} films was achieved with the application of a thin \ce{SiO2} capping layer utilizing growth conditions typical of the GaAs/\ce{SiO2}  system. The relationship between the thickness of \ce{SiO2} caps and growth selectivity was examined, revealing that sub-1 nm capping layers can significantly influence the mask surface  chemistry, indicating that by depositing a thin layer of \ce{SiO2}, \ce{SiO2}-like selectivity for any mask material can be realized without degrading its optical response.

\end{abstract}

\section{Introduction}
Selective-area embedded regrowth of III–V semiconductors by molecular beam epitaxy (MBE) enables the seamless integration of metals, dielectrics, and crystalline semiconductors. By defining crystal seeding regions with a mask and then encapsulating the amorphous structure in single crystalline III-V, it creates a unique design space for the development of optoelectronic and photonics devices. Deterministic growth of low-dimensional quantum emitters \cite{Schumann_2011,Desplanque_2018,Birudavolu2004} and all-epitaxial photonic crystals,\cite{Noda2017,Gelleta2015,Nelson2011,Cho2011,Huang} embedded high-contrast photonics,\cite{Wang2017,LiuJZ2014} and laterally-structured plasmonic layers\cite{Skipper2022} are promising avenues enabled by this all-MBE approach to selective embedded regrowth.
%One example of applications that would benefit from the addition of alternate materials to the capabilities of MBE SAG is embedded long-wave infrared high contrast photonics.

The predominant mask materials in selective-area growth (SAG) are \ce{SiO2} and \ce{Si3N4}, due to their inert nature, however relying on these masks alone for design of SAG-based devices limits their potential. For instance, these well-established masks have high extinction coefficients in the 8 $\mu$m to 12 $\mu$m wavelength range, prohibiting the design of embedded high contrast photonics in the long-wave infrared due to high losses. Similarly, the development of embedded metasurfaces would benefit from the use of different mask materials with a diverse range of refractive indices, both higher and lower than the host III-V semiconductor, coupled with low extinction coefficients. Therefore, there is a need to expand the capabilities of selective-area MBE and enable access to a greater diversity of compatible mask materials. However, all-MBE selective-area regrowth has been historically difficult to achieve because of its poor III–V deposition selectivity, which leads to an aggregation of adatoms on the mask surface and ultimately the parasitic formation of polycrystalline material.\cite{Allegretti1995,Aseev2019,Horikoshi1999,Yokoyama1989,Sugaya_1992,Lee2002} Therefore, selective regrowth of technologically-interesting materials that may not have the same inert quality as \ce{SiO2} or \ce{Si3N4}  may be more difficult to regrow, requiring potentially more extreme conditions outside of the ideal growth regime. As a result, there have been few III-V MBE SAG demonstrations using alternate dielectric masks. 

Here, we explore the III-V solid-source (SS-)MBE deposition selectivity of \ce{Al2O3}, \ce{TiO2}, and \ce{HfO2} with respect to well-established mask materials. These oxides serve as essential dielectric and interface materials due to their ultra-high vacuum stability, insulating behavior, and inert chemical properties, and therefore have applications in photonics and optoelectronics (i.e. optical waveguide, dielectric mirror) or in hybrid quantum devices.\cite{Heindel_2012_tioxgmr,DUBEY2017tiosio2bragg,selective_tejedor_2019,JiangPbTeHfO22022,mechanism_kishino_2019, effect_nagae_2011, improved_kishino_2009} Despite the promise, there have been few III-V MBE SAG demonstrations using these mask materials. Tejedor et al.\cite{selective_tejedor_2019} first employed \ce{HfO2} for III-V SAG as a high-$\kappa$ gate dielectric selective-area mask of InGaAs nanowires toward highly scaled nMOS devices in 2019, however, its use has seen little progress since the initial demonstration. On the other hand, \ce{TiO2} and \ce{Al2O3} have been used widely in selective area MBE of nitrides for low-defect density growth, nanowires, micro-light-emitting diodes, and photonic resonators\cite{mechanism_kishino_2019, effect_nagae_2011, improved_kishino_2009}, but \ce{Al2O3} has little exploration and \ce{TiO2} has not been explored for III-V SS-MBE SAG. Outside of the materials of interest, some group III oxides and III-V native oxides have also been investigated by III-V MBE and chemical beam epitaxy SAG, such as GaAs/AlGaAs native oxides, and \ce{In2O3}, however, like \ce{HfO2}, there has been little subsequent use.\cite{effect_matsuda_2011,deposition_ozasa_1994,area_yoshiba_2007} It is evident that there is a need to both derive a greater understanding of III-V/mask growth selectivity and find a path forward to incorporating new mask materials into single crystalline III-V, even those that may be highly non-selective or even reactive. Expanding the library of III-V embedded oxides could realize advanced capabilities for next-generation (opto)electronics and photonics.

In this study, we first surveyed the selective growth properties of the emerging mask materials using a study of polycrystalline GaAs surface coverage after deposition on unpatterned films and demonstrated that \ce{Al2O3} has promising selective growth characteristics within the typical GaAs growth temperature regime. However, it also was observed that \ce{HfO2} was highly non-selective and \ce{TiO2} was reactive with Ga deposition. Surface modification was proposed as a technique to alter selective growth characteristics and then was demonstrated on \ce{Si3N4} and \ce{TiO2} films by adding a thin layer of \ce{SiO2} on top of the film via plasma-enhanced chemical vapor deposition; no polycrystalline GaAs deposition was observed under selective growth conditions that produce polycrystalline material on the unmodified films. The influence of the thickness of the capping layer was then explored with the \ce{Al2O3} system and it was shown that while the selectivity did change linearly between thicknesses of $\sim 1$ to 10 nm, there was a significant difference between the poly-GaAs characteristics on bare \ce{Al2O3} and \ce{Al2O3} with a sub-nanometer cap of \ce{SiO2}, effectively decoupling the selective growth properties from the optical response of the film by modifying its the surface chemistry.

\section{Experimental Section}
\subsection{MBE Compatibility}

The emerging mask materials in this study were identified due to their broad range of optical constants\cite{Kischkat:12,Bright2012OpticalPO} and their compatibility with the molecular beam epitaxy system. Explored materials were required to be inorganic to avoid carbon contamination and the critical properties of the materials that we identified were vapor pressure and sublimation temperature. Growth of III-V semiconductor via MBE requires substrate temperatures as high as 700 to 750 $^\circ$C, as measured by bandedge thermometry. Therefore, the mask materials needed to be stable in this temperature range such that the sublimation point was far above this regime and the vapor pressure remained sufficiently low. Due to the success of \ce{SiO2}, the properties of this thin film were used a metric for the chosen mask materials; all vapor pressures and sublimation temperatures were comparable to \ce{SiO2}. \cite{Schulz2005ReviewOA,Ryklis1969}

\subsection{Film Fabrication}

\ce{SiO2} and \ce{Si3N4} films were deposited via plasma-enhanced chemical vapor deposition (PECVD) using a PlasmaTherm 790. In order to prepare the PECVD chamber for UHV-compatible fabrication, the system was first cleaned and a $\sim$250 nm layer of \ce{SiO2} was deposited on the chuck to create an ultraclean surface and outgas the precursor lines. Prior to deposition, the native oxide of the GaAs wafer was stripped with a 1 minute 37$\%$ hydrochloric acid etch and then a 2 minute DI water rinse and confirmed by the resulting hydrophobicity of the wafer. It was then inserted onto the chuck for deposition of the film. After PECVD of the \ce{SiO2} film on GaAs, \ce{Al2O3}, \ce{HfO2} and \ce{TiO2} were deposited with atomic layer deposition (ALD). Following ALD, a 30 min ultraviolet (UV) Ozone irradiation process is performed to remove residual carbon from the film surface that may result from the use of organic precursors.

\subsection{Molecular Beam Epitaxial Growth}

After film preparation, the samples were baked at 450 $^\circ$C, measured by thermocouple, in the buffer chamber under an atomic hydrogen flux supplied using a Veeco Atomic Hydrogen source (AHS) equipped with an inline Pd purifier. Samples were then transferred into the growth chamber and baked at 450 $^\circ$C oriented away from the cells for 10 min. After rotating the sample toward the sources, co-deposition of Ga and \ce{As4} was performed on the various films using a Veeco SUMO effusion cell for Ga and a Veeco Mark IV Arsenic cracker with the cracking zone set to 650 $^\circ$C. A growth rate of 0.25 $\mu$m/hr under a 12.8 \ce{As4}/Ga flux ratio was used for all growths in this study. The total deposition time was 23.7 min, which is the equivalent of 100 nm of GaAs growth under unity sticking conditions.

\subsection{Sample Characterization}

Polycrystalline deposition on films were imaged using scanning electron microscopy (Neon 40 and Gemini 460 systems). MATLAB processing techniques discussed in the supplemental information of our previous work were utilized to determine nuclei morphology and characteristics.\cite{garcia_experimentally_2025} Atomic force microscopy of the films was performed using a Bruker Dimension Icon scanning probe microscope and a VersaProbe4 X-ray Photoelectron Spectrometer was used to characterize the surface composition of the prepared films. %Fourier transform infrared spectroscopy was performed with a Bruker Vertex 80v FTIR with a Bruker Hyperion Microscope. 

\begin{figure}[H]
	\centering
	\includegraphics[page=1,scale=1,clip,trim=0in 3.83in 3.6in 0in]{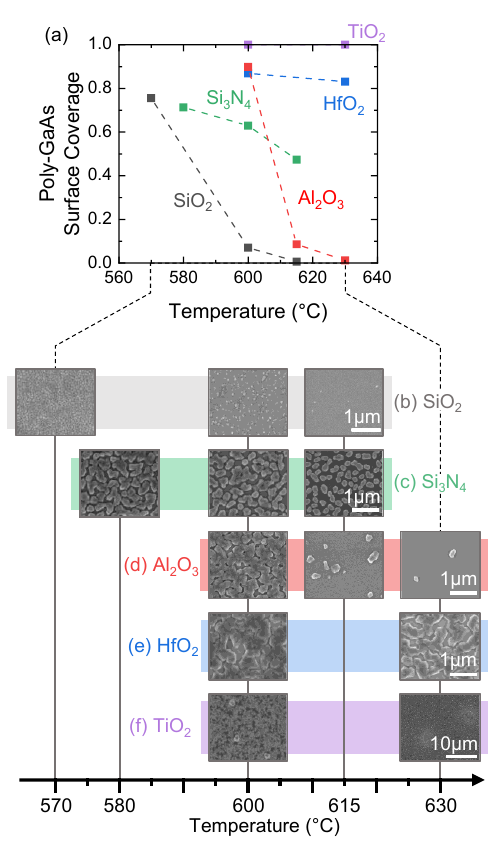}
	\caption{(a) Graph of resulting polycrystalline surface coverage values and (b)-(f) representative planview SEM images for each film after deposition of 100 nm of GaAs on unpatterned films after different substrate temperatures show relative mask selectivity.}
	\label{fig:SCAll}
\end{figure}

\section{Results and Discussion}

\subsection{Selectivity Survey}

An exploratory study of material selectivity was performed by observing polycrystalline GaAs growth on unpatterned films under identical MBE conditions. Co-deposition of Ga and As was performed on the unpatterned films at a rate of 0.25 $\mu$m/h under a high tetrameric arsenic overpressure with a flux ratio of 12.8 at substrate temperatures ranging from 570 to 630 $^\circ C$ to span the ideal growth regime and typical selective growth temperatures for GaAs. After 100 nm GaAs deposition on the various films, planview scanning electron microscopy (SEM) observed variations in the amount of impinged GaAs and resulting nuclei morphology as shown in Figure~\ref{fig:SCAll}b. The observed characteristics were correlated with the magnitude of the selective growth mechanisms (Ga adsorption, desorption, diffusion) providing information regarding how readily Ga impinges on the non-crystalline surface and its mobility.  Surface coverage, the fractional coverage of the polycrystalline III-V material on film surface, combines these attributes into a single metric to describe growth selectivity as a lower surface coverage can be correlated with increased desorption, increased diffusion, and decreased adsorption.

Figure~\ref{fig:SCAll}a shows the relative selectivity of the materials of interest with respect to the substrate temperature. \ce{Al2O3} performed the best of the alternative mask materials as its surface coverage approached a near-zero value at the upper end of the GaAs growth temperature range, indicating increased adatom mobility during growth that may be sufficient to achieve selectivity at 630 $^\circ$C on patterned templates. The surface coverage trend showed a similar temperature relationship as compared to \ce{SiO2} indicating that their selectivity may be driven by the same limiting mechanism, which was hypothesized to be gallium bonding with oxygen vacancies. Despite the similarity in the slope of the trend, a higher temperature growth regime was required for near-zero surface coverage under these conditions due to increased adsorption on the mask surface that was likely a result of the preference Ga has for surfaces containing group III atoms at these substrate temperatures. The nuclei that formed on the \ce{Al2O3} surface at 630 $^\circ$C were $\sim$3$\times$ larger in nuclei radius than \ce{SiO2} at the same temperature and increased average nuclei or polycrystalline grain radii were observed for equivalent surface coverage values, potentially indicating a higher diffusion length of the impinged gallium adatoms.

Contrary to \ce{SiO2} and \ce{Al2O3}, the silicon nitride mask exhibited a lower temperature sensitivity; this may be attributed to the greater stability of the Ga-N bond at this range of temperatures as compared to the Ga-O bond.\cite{ASAOKA200340,Ambacher} An increase in nuclei radii was also observed; the average in-plane nuclei radius at 615 $^\circ$C of poly-GaAs on \ce{Si3N4} was approximately $\sim$3$\times$ that of the nuclei on \ce{SiO2} which can be attributed likely to a higher adsorption coefficient. It is evident based on these aspects of the nuclei morphology that selective growth under these conditions would require a higher minimum temperature or a lower growth rate than can be achieved with the \ce{SiO2}.

A near-zero surface coverage regime was not observed for the other two films. The \ce{HfO2} film showed a high temperature nuclei stability; up to 630 $^\circ$C the surface coverage did not decrease below 0.8 indicating that adsorption of gallium adatoms was likely the dominating mechanism. During the 600 to 630 $^\circ$C growths, reflection high energy electron diffraction (RHEED) measurements observed rings indicative of polycrystalline growth after just 2 to 3 nm of GaAs deposition. Increasing the substrate temperature to 650 $^\circ$C saw a very minor delay in the arrival of polycrystalline rings in RHEED to 5 nm of deposition indicating that there may be a path toward selectivity combining temperatures above 650 $^\circ$C with lower growth rates. While this may yield selectivity, it is not viable for regrowth of III-Vs as increasing the temperature to $\sim$100 $^\circ$C above the ideal growth temperature could result in entropic roughening or inhibit growth in the window regions all together.

The resulting nuclei morphology on the \ce{TiO2} film was disparate as compared with all the other films. The formation of multiple types of nuclei at the different growth temperatures is potentially indicative of interfacial reactions between the incident adatoms and the \ce{TiO2} film\cite{Schmid-Fetzer1988}. Furthermore, during the growth, a rapid apparent temperature increase of $\sim$50 $^\circ$C was observed in the band-edge thermometry measurement starting immediately after opening the gallium shutter; this temperature change could result from reactions Ga reacting with the surface or an apparent roughening of the film interface due to this atypical formation of polycrystalline material. Prior to growing on these films, they were annealed under an atomic hydrogen source, which has been shown to hydrogenate \ce{TiO2} films, modifying the Ti oxidation state and introducing oxygen vacancies, which could result in a more reactive dielectric surface. Due to the atypical behavior of this material during the growth survey, a direct comparison in growth selectivity was not made, however it is evident that is not compatible with III-V SAG as-is and would require either modifications to the MBE process or film preparation.

\begin{figure}[H]
	\centering
	\includegraphics[page=2,scale=1,clip,trim=0in 6.6in 3.58in 0in]{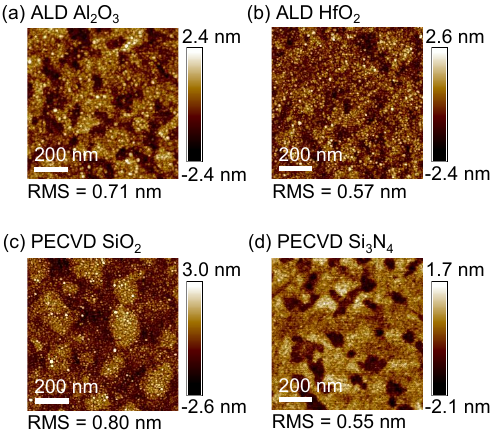}
	\caption{Atomic force microscopy of the films prior deposition to evaluate surface morphology. The roughness of all emerging films was observed to comparable to \ce{SiO2} indicating that the degradation in selectivity was not a result of poor surface quality.}
	\label{fig:AFMall}
\end{figure}

The surface roughness of \ce{Al2O3}, \ce{HfO2}, \ce{SiO2} and \ce{Si3N4} was first characterized to understand if it was driving the differences in deposition selectivity. All films were observed to have sub-nanometer root-mean-square roughness (RMS) as shown in Figure~\ref{fig:AFMall} and the RMS values were not correlated to the amount of deposited polycrystalline GaAs or relative selectivity of the film. In fact, \ce{SiO2} had the highest RMS roughness despite it being the most selective mask; it is evident that the composition of the mask surface is more likely the driving factor in this case than the morphology of the surface.

\begin{figure}[H]
	\centering
	\includegraphics[page=3,scale=1,clip,trim=0in 3.5in 3.56in 0in]{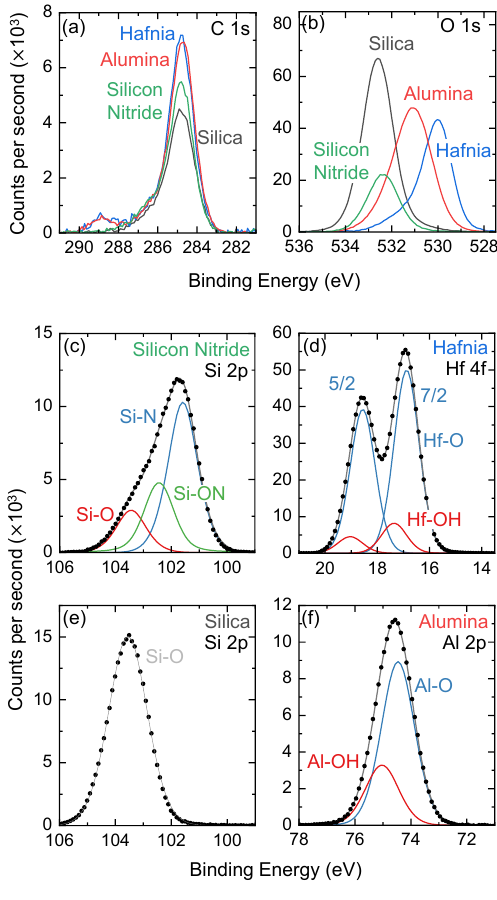}
	\caption{Graphs comparing the (a) O 1s and (b) C 1s XPS spectra of the dielectric films.  The surface composition calculate from the XPS measurements is displayed in a table. The data showed a correlation between decreased O 1s binding energy and increased carbon content with a decrease in growth selectivity. XPS spectra of the characteristic peaks of the dielectric films (c) silicon nitride, (d) hafnia, (e) silica and  (f) alumina. Complete XPS data can be found in the Supplemental Information.}
	\label{fig:XPS}
\end{figure}

X-ray photoelectron spectroscopy (XPS) was performed on the films to probe compositional differences and bonding states. Figure \ref{fig:XPS} shows a symmetric peaks in the O 1s and Si 2p spectra representative of \ce{SiO2}. Initial observations showed that the lower binding energy of the O 1s orbital was correlated with increased GaAs deposition at 600 $^\circ$C indicating that oxygen sites on the surface may be playing a role in the limiting mechanism of selectivity. Evidence of oxygen vacancies was present in both the \ce{HfO2} and \ce{Al2O3} spectra.

The Hf 4f, O 1s and C 1s spectra were measured on the \ce{HfO2}. As seen in Figure~\ref{fig:XPS}b, the Hf 4f spectra showed the two characteristic peaks of the spin-orbit doublet corresponding to the 4f$_{5/2}$ and 4f$_{7/2}$ orbitals at 18.5 eV and 16.9 eV respectively with the expected intensity ratio of $\sim$0.75 and peak splitting of $\sim$1.6 eV. A spectral contribution to the doublet from suboxide states was observed, indicating the presence of poorly oxidized hafnium and oxygen vacancies.\cite{MARTINEZPUENTE2022115964,Triyoso_2004,GRITSENKO20161} This is consistent with the shape of the O 1s spectra, which exhibits an asymmetry with a tail extending toward higher binding energies representative of Hf-OH groups.\cite{Triyoso_2004} %While this tail may also be indicative of Hf-silicate group, it is unlikely that the measurement is probing the \ce{HfO2}/\ce{SiO2} interface due to the limited penetration depth of the technique nor do we expect the presence of Si on the film surface due to the low temperature nature of the ALD process.\cite{SAMMELSELG2007150,Li_2020} 

A similar asymmetry was observed in the \ce{Al2O3} spectra. The analyzed spectra for this film type were the Al 2p, O 1s and C 1s orbital states. The Al 2p spectra and O 1s were found at 74.6 eV and 531 eV respectively, which are consistent with known values of the Al-O bond.\cite{Balme_2015, Broas_2017,Ozaki_2023} Both exhibit an asymmetric peak with a higher energy tail indicating the presence of Al-OH groups, and therefore a more reactive surface than stoichiometric \ce{Al2O3}.\cite{Broas_2017} The use of \ce{H2O} as a source of oxygen for ALD has been shown to produce films with a higher OH$^-$ content, therefore, an alternate source of oxygen may enable the reduction of these groups and increased deposition selectivity.\cite{Ozaki_2023}

In the \ce{Si3N4} film, a similar surface reactivity argument can be made for the degradation in selectivity. The Si 2p, N 1s, O 1s and C 1s states were measured using XPS. An asymmetric peak with a higher energy tail was observed for the Si 2p orbital. Deconvolution of the peaks show the presence of \ce{SiO2} and \ce{SiON} groups on the surface of the \ce{Si3N4} film in addition to the main nitride peak that is likely due to a combination of deposition conditions and surface oxidation. The non-stoichiometric SiO$_x$N$_y$ is likely to have dangling bonds on the surface that can act as bonding sites for Ga and is consistent with the hypothesis that available nitrogen and oxygen bonding sites degrade growth selectivity.

 Similarly, a trend in carbon content on the surface was also observed. A greater carbon content was observed for films with high poly-GaAs deposition rates. The differences in carbon content may be a result of different processing conditions used to fabricated the films. Carbon contamination could potentially result in low-energy nucleation sites increasing the rate of nuclei formation. The path forward for mitigating this potential influence on selectivity is optimization of the ALD and PECVD conditions to minimize the reactivity of the surface when exposed to III-V MBE deposition.

\begin{figure}[H]
	\centering
	\includegraphics[page=4,scale=1,clip,trim=0in 5.65in 3.58in 0in]{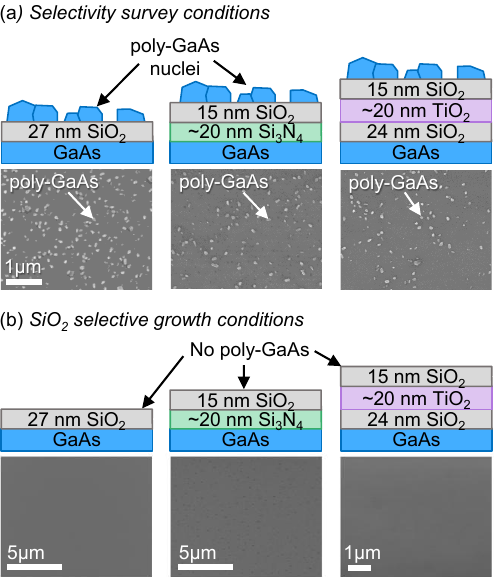}
	\caption{(a) Planview SEM images of films after 100 nm of GaAs deposition at 600 $^\circ$C on the \ce{Si3N4} and \ce{TiO2} films after being modified with a 15 nm \ce{SiO2} cap show nearly identical nuclei morphology to that of \ce{SiO2} film after surface modification. (b) After 100 nm of GaAs deposition on the modified \ce{Si3N4} and \ce{TiO2} films, planview SEM images show that no polycrystalline material was deposited, identical to the \ce{SiO2} film, indicating the ability to access the selective growth regime of $\ce{SiO2}$ with this surface modification.}
	\label{fig:Proof-of-concept}
\end{figure}

\subsection{Surface Modification}
Despite the furthered understanding of mask selectivity, the high reactivity of films such as \ce{HfO2} and \ce{TiO2} remained a significant barrier to their use in selective growth applications. In performing the selectivity survey, it was demonstrated that the addition of \ce{Al2O3}, \ce{HfO2} and \ce{TiO2}  film on top of \ce{SiO2} degraded growth selectivity, therefore, it was hypothesized that the reverse could be leveraged as a "surface modification" to enable selective MBE regrowth of highly non-selective or reactive materials using growth conditions suitable for \ce{SiO2} films.

The initial proof-of-concept was performed by including a 15 nm capping layer of \ce{SiO2} on the \ce{Si3N4} and \ce{TiO2} films with PECVD. The same aforementioned survey growth conditions were utilized and samples were regrown at the 600 $^\circ$C substrate temperature. The introduction of the capping layer enabled nuclei morphology and surface coverage values nearly identical to the \ce{SiO2} film control growth (Figure~\ref{fig:Proof-of-concept}a).  By then incorporating a $10\%$ periodic supply epitaxy cycle (6s Ga shutter open, 54s closed), the selective growth regime of \ce{SiO2} was able to be leveraged to completely mitigate polycrystalline material formation on the capped \ce{Si3N4} and \ce{TiO2} films under conditions that would otherwise have not been selective. It's highly beneficial to be able to modify materials such that they can be selectively regrown within the \ce{SiO2} regime as its inert nature currently allows for growth closest to that of the ideal III-V regime, which is necessary for high optical quality and band-engineering. 

\begin{figure}[t]
	\centering
	\includegraphics[page=5,scale=1,clip,trim=0.1in 5.74in 0in 0in]{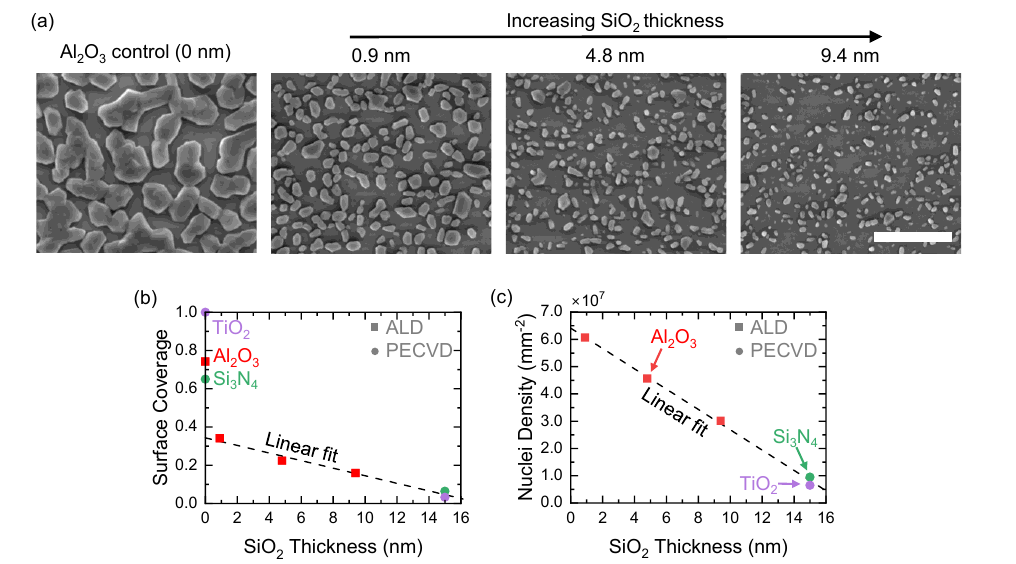}
	\caption{(a) Planview SEM images of films after 100 nm of GaAs deposition at 600 $^\circ$C on the ALD SiO$_x$ modified \ce{Al2O3} unpatterned films with an \ce{Al2O3} control showing changing nuclei morphology with varying SiO$_x$ thickness. Graphs of (b) polycrystalline surface coverage and (c) nuclei density show approximately linear relationships with respect to the thickness of the SiO$_x$ modification that approaches the selectivity properties of PECVD \ce{SiO2} as the thickness increases.}
	\label{fig:SiOonAlO}
\end{figure}

This is significant for embedded photonics and metasurfaces, however it is necessary to limit the influence of the capping layer on the spectral response when designing embedded structures, therefore efforts were taken to minimize the thickness of the modifying layer to achieve a traditional surface modification approach. The ideal case would be the use of a single treatment step to passivate the weak bonding sites on the film surface with the more preferable Si-O groups to achieve \ce{SiO2}-like growth selectivity. ALD of \ce{SiO2} was performed on \ce{Al2O3} films identical to those from the previously discussed selectivity survey study. The thickness of this \ce{SiO2} modification layer was varied from 0.9 nm to 9.4 nm. At a substrate temperature of 600 $^\circ$C, 100 nm of GaAs deposition was performed on the films. There was a significant decrease in polycrystalline GaAs build-up on the film with just 0.9 nm of the \ce{SiO2} modification, after which, the measured surface coverage and nuclei density of the resulting polycrystalline films followed an approximately linear relationship for the SiO$_x$-modified films approaching the benchmark PECVD \ce{SiO2} values. The increasing coverage of the Si-O bonds as the thickness of the modification layer increases results in a less reactive surface and improved selectivity (Figure~\ref{fig:SiOonAlO}). 

\section{Conclusion}
In conclusion, this study underscores the potential of selective-area embedded regrowth of III–V semiconductors via MBE to enable innovative photonics and optoelectronic devices through the integration of diverse mask materials. While traditional mask materials such as \ce{SiO2} and \ce{Si3N4} pose limitations in technologically-relevant wavelength ranges, the exploration of alternative materials such as \ce{Al2O3}, \ce{TiO2}, and \ce{HfO2} reveals promising opportunities for expanding selective regrowth capabilities. Our findings indicate that \ce{Al2O3} offers favorable selectivity within the typical GaAs growth temperature regime, although \ce{SiO2} currently remains superior in terms of deposition selectivity. \ce{HfO2} and \ce{TiO2} on the other hand were not selective under typical regrowth conditions, therefore, we introduced the use of surface modification techniques for improving selective area MBE growth characteristics of a mask material. By applying a thin \ce{SiO2} capping layer to the mask, highly non-selective or reactive mask materials can be made inert, mitigating the polycrystalline deposition on less selective materials and enabling the use of the well-established III-V/\ce{SiO2} growth regimes. This approach opens new avenues for leveraging different mask materials to enhance growth selectivity without compromising on growth selectivity or optical performance.

\begin{acknowledgement}

This work was supported by the National Aeronautics and Space Administration (NASA) through the Quantum Pathways Institute (Award 80NSSC22K0287). 

\end{acknowledgement}


\bibliography{achemso-demo}
\newpage
%\noindent \textbf{For Table of Contents Use Only}

%\noindent Title: Surface Modification for III-V Selective Area Molecular Beam Epitaxy of Non-Selective Mask Materials

%\noindent Authors: Ashlee M. Garc\'ia, Byron D. Aguilar, William J. Doyle, Pernille Undrum Fathi, Federico Capasso, Daniel Wasserman, and Seth R. Bank

%\noindent TOC Graphic:
%\newline

%\includegraphics[page=6,scale=1,clip,trim=0in 7.84in 3.64in 0in]{260410_SurfaceModification_allFigures-combined.pdf}

%\noindent Synopsis: We explore emerging mask materials for III-V selective area growth by molecular beam epitaxy and demonstrate a new surface modification approach to realize regrowth of highly non-selective dielectrics for advanced design of all-epitaxial photonic and quantum devices. 

\noindent\textbf{Supplemental Information}

\vspace{1em}
\noindent\textit{X-ray Photoelectron Spectroscopy}

X-ray photoelectron spectroscopy (XPS) was performed at the Texas Materials Institute (TMI) using a VersaProbe4 instrument equipped with an Al source. The characteristic peaks of each film were measured in addition to carbon. The spectra were aligned according to the known position of the C 1s photoelectron peak. After which,  multi-peak fitting of the measured data was performed with CasaXPS using a Shirley background and Gaussian-Lorentzian product line-shape. Figures 1-4 below present the measured data of the characteristic peaks of all films in this study along with the peak fit.

\vspace{1em}
\noindent\textit{Residual Gas Analysis}

Residual gas analysis of the molecular beam epitaxy (MBE) growth chamber was used to confirm that no contamination occurred as a result of growth on these films deposited by ALD. A 200 amu Stanford Research Systems (SRS) residual gas analyzer (RGA) equipped with an electron multiplier was used to perform this measurement. Figure 5 below shows the before and after scans of the chamber on the first introduction of these films into the chamber.

\setcounter{figure}{0}

\begin{figure}[H]
	\centering
	\includegraphics[page=23,scale=.95,clip,trim=0in 5.46in 2.45in 0in]{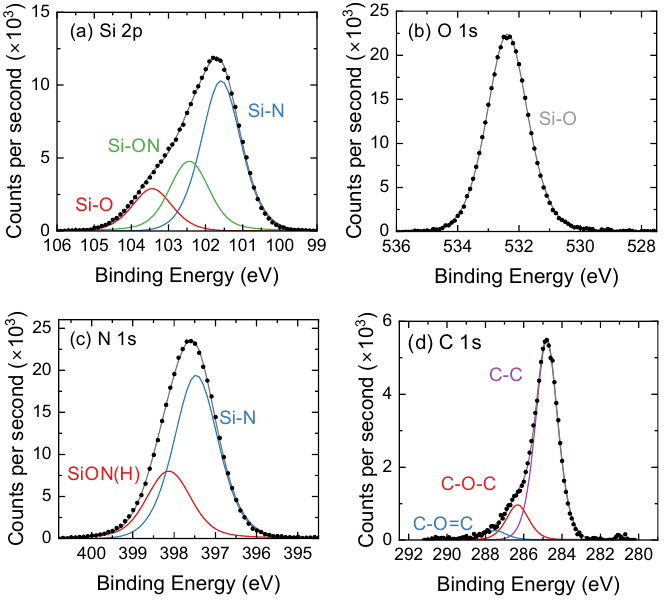}
	\caption{XPS spectra of \ce{Si3N4} film showcasing the (a) Si 2p, (b) O 1s, (c) N 1s and (d) C 1s peaks. Measured data is represented with black circles and the solid lines represent the fitted peaks.$^{1-3}$ }
	\label{fig:RGATiO21}
\end{figure}

\begin{figure}[H]
	\centering
	\includegraphics[page=23,scale=.95,clip,trim=0in 3.04in 0.1in 4.5in]{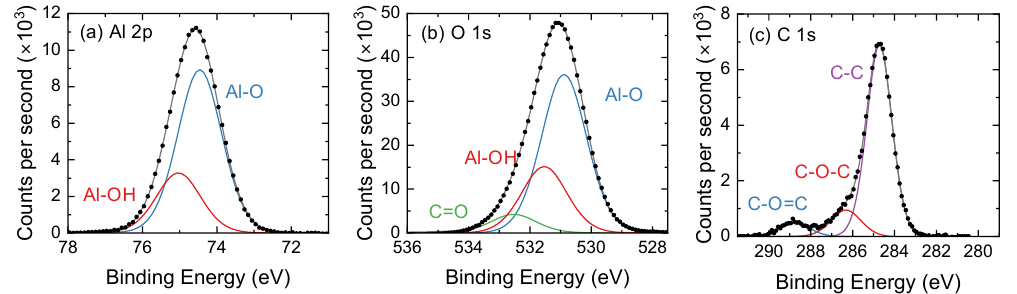}
	\caption{XPS spectra of \ce{Al2O3} film showcasing the (a) Al 2p, (b) O 1s and (c) C 1s peaks. Measured data is represented with black circles and the solid lines represent the fitted peaks.$^{4,5}$}
	\label{fig:RGATiOfirst}
\end{figure}

\begin{figure}[H]
	\centering
	\includegraphics[page=23,scale=.95,clip,trim=0in 0.57in 0.1in 6.88in]{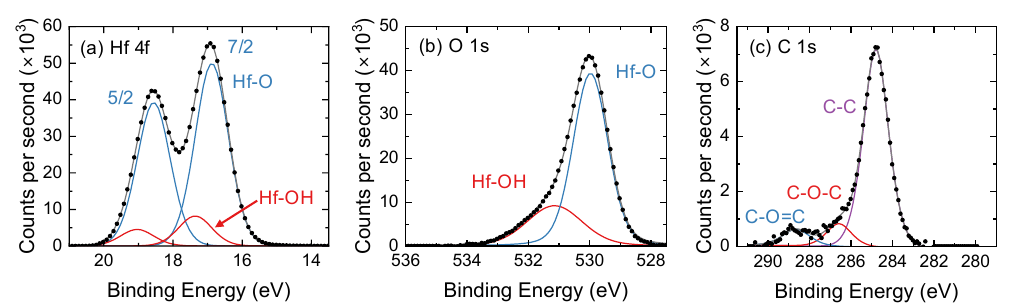}
	\caption{Graph of the XPS spectra of \ce{HfO2} film showcasing the (a) Hf 4f, (b) O 1s and (c) C 1s peaks. Measured data is represented with black circles and the solid lines represent the fitted peaks.$^{6,7}$}
	\label{fig:RGATiO2}
\end{figure}

\begin{figure}[H]
	\centering
	\includegraphics[page=24,scale=.95,clip,trim=0in 7.5in 0.1in 0in]{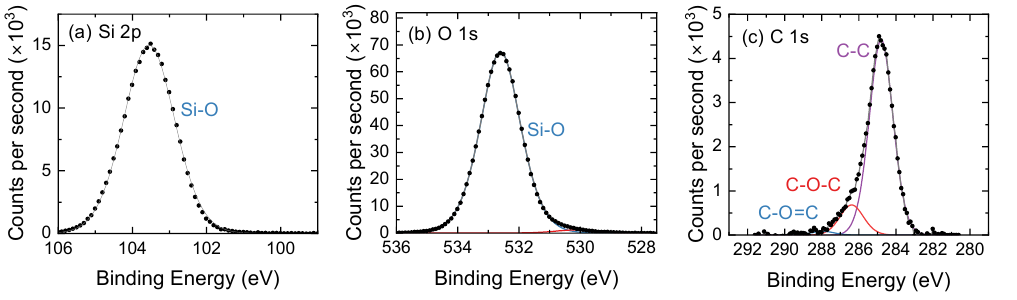}
	\caption{XPS spectra of \ce{SiO2} film showcasing the (a) Si 2p, (b) O 1s and (c) C 1s peaks. Measured data is represented with black circles and the solid lines represent the fitted peaks.$^{1,2}$}
	\label{fig:RGATiO22}
\end{figure}

\begin{figure}[H]
	\centering
	\includegraphics[page=22,scale=.95,clip,trim=0in 1.45in 1.61in 0in]{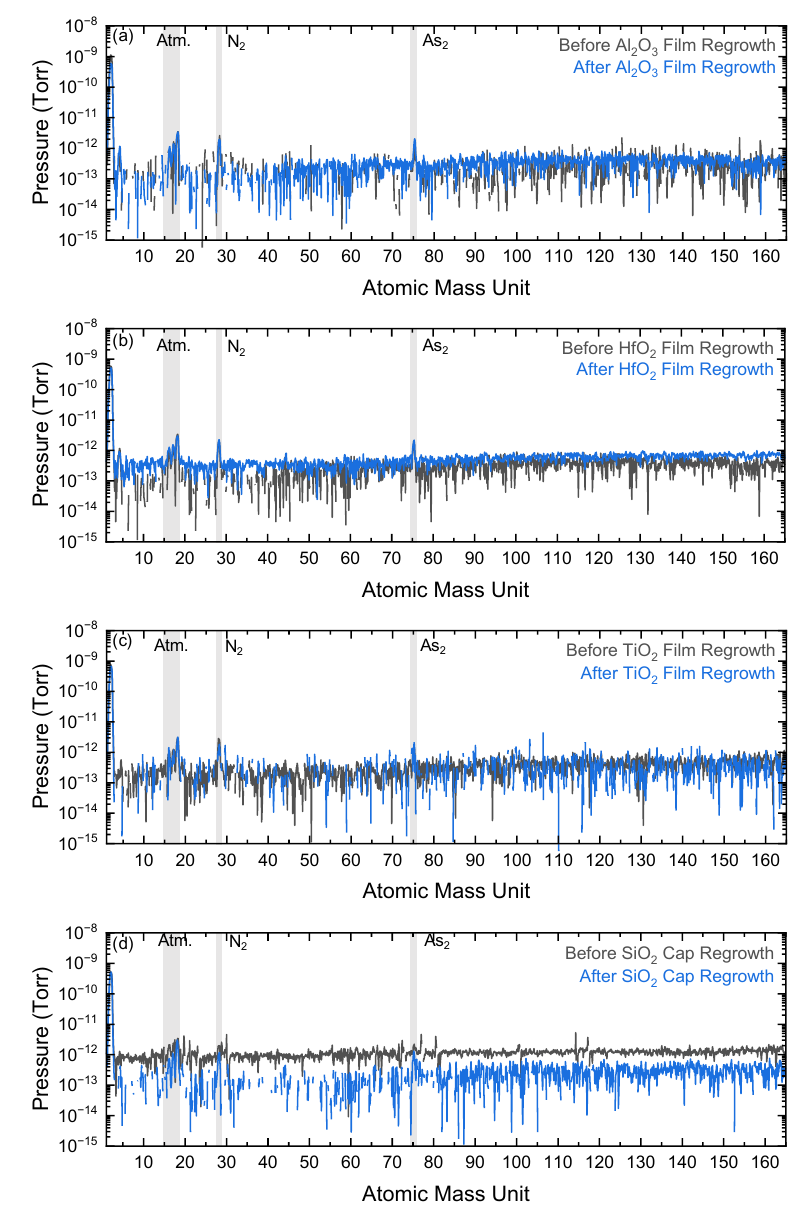}
	\caption{Residual gas analysis of MBE growth chamber before and after selectivity survey growth was performed on (a) \ce{Al2O3}, (b) \ce{HfO2}, (c) \ce{TiO2}, and (d) \ce{SiO2}-capped \ce{Si3N4} and \ce{TiO2} films. Scans shows no contamination or change in the growth chamber environment.}
	\label{fig:RGASiO2cap}
\end{figure}

\noindent\textbf{References}
\begin{enumerate}[label=(\arabic*)]
    \item Lukose, R.; Lisker, M.; Akhtar, F.; Fraschke, M.; Grabolla, T.; Mai, A.; Lukosius, M. Influence of plasma treatment on SiO2/Si and Si3N4/Si substrates for large-scale transfer of graphene. \textit{Scientific Reports} \textbf{2021}, 11, 13111.
    \item Viard, J.; Beche, E.; Perarnau, D.; Berjoan, R.; Durand, J. XPS and FTIR study of silicon oxynitride thin films. \textit{Journal of The European Ceramic Society,} \textbf{1996}.
    \item Mello-Castanho, S. R. H. d.; Moreno, R.; Fierro, J. Influence of process conditions on the surface oxidation of silicon nitride green compacts. \textit{Journal of Materials Science}, \textbf{1996}.
    \item Balme, S.; Iatsunskyi, I.; Iatsunskyi, I.; Kempi´nski, M.; Kempi´nski, M.; Jancelewicz, M.;
Jancelewicz, M.; Za leski, K.; Za leski, K.; Jurga, S.; Jurga, S.; Smyntyna, V.; Smyntyna,
V. Structural and XPS characterization of ALD Al2O3 coated porous silicon.
Vacuum 2015,
\item Iatsunskyi, I.; Kempi´nski, M.; Jancelewicz, M.; Za leski, K.; Jurga, S.; Smyntyna, V.
Structural and XPS characterization of ALD Al2O3 coated porous silicon. \textit{Vacuum} \textbf{2015},
113, 52–58.
\item Mart\'inez-Puente, M.; Horley, P.; Aguirre-Tostado, F.; L\'opez-Medina, J.; Borb\'on-
Nu$\tilde{n}$ez, H.; Tiznado, H.; Susarrey-Arce, A.; Mart\'inez-Guerra, E. ALD and PEALD
deposition of HfO2 and its effects on the nature of oxygen vacancies. \textit{Materials Science
and Engineering: B} \textbf{2022}, 285, 115964.
\item Triyoso, D. H. et al. Impact of Deposition and Annealing Temperature on Material and
Electrical Characteristics of ALD HfO2. \textit{Journal of The Electrochemical Society,} \textbf{2004}.
\end{enumerate}

\end{document}